\documentclass[1p,times]{elsarticle}
\usepackage{graphicx}
\usepackage{ecrc}

\volume{00}

\firstpage{1}

\journalname{Physics Procedia}

\runauth{M.\ Bachmann}

\jid{phpro}

\jnltitlelogo{Physics Procedia}

\usepackage{amssymb}
\biboptions{compress}

\begin{document}

\begin{frontmatter}

%% Title, authors and addresses

%% use the tnoteref command within \title for footnotes;
%% use the tnotetext command for the associated footnote;
%% use the fnref command within \author or \address for footnotes;
%% use the fntext command for the associated footnote;
%% use the corref command within \author for corresponding author footnotes;
%% use the cortext command for the associated footnote;
%% use the ead command for the email address,
%% and the form \ead[url] for the home page:
%%
%% \title{Title\tnoteref{label1}}
%% \tnotetext[label1]{}
%% \author{Name\corref{cor1}\fnref{label2}}
%% \ead{email address}
%% \ead[url]{home page}
%% \fntext[label2]{}
%% \cortext[cor1]{}
%% \address{Address\fnref{label3}}
%% \fntext[label3]{}

\dochead{}
%% Use \dochead if there is an article header, e.g. \dochead{Short communication}

\title{Contact-density analysis of lattice polymer adsorption transitions}

%% use optional labels to link authors explicitly to addresses:
%% \author[label1,label2]{<author name>}
%% \address[label1]{<address>}
%% \address[label2]{<address>}

\author{Michael Bachmann}
\ead{bachmann@smsyslab.org}
\ead[url]{http://www.smsyslab.org}

\address{Soft Matter Systems Research Group, Institut f\"ur Festk\"orperforschung (IFF-2),\\
Forschungszentrum J\"ulich,
D-52425 J\"ulich, Germany}
\begin{abstract}
By means of contact-density chain-growth simulations, 
we investigate a simple lattice model of a flexible polymer interacting with an attractive substrate. 
The contact density is a function of the 
numbers of monomer--substrate and monomer--monomer contacts. These contact numbers represent
natural order parameters and allow for a comprising statistical study of the conformational space accessible
to the polymer in dependence of external parameters such as the attraction strength of the substrate
and the temperature. Since the contact density is independent of the energy scales associated to the
interactions, its logarithm is an unbiased measure for the entropy of the conformational space.
By setting explicit energy scales, the thus defined, highly general microcontact entropy
can easily be related to the microcanonical entropy 
of the corresponding hybrid polymer--substrate system.
\end{abstract}

\begin{keyword}
%% keywords here, in the form: keyword \sep keyword

%% MSC codes here, in the form: \MSC code \sep code
%% or \MSC[2008] code \sep code (2000 is the default)
microcanonical analysis \sep adsorption \sep small system \sep polymer \sep chain-growth simulation
\end{keyword}

\end{frontmatter}

%%
%% Start line numbering here if you want
%%
% \linenumbers

%% main text
\section{Introduction}
All system-relevant informations are encoded in the density of states $g(E)$ which can be 
understood as the volume of the phase space associated to a certain system energy $E$. For
a discrete system, this corresponds to the number of microstates with the same energy $E$, i.e.,
it represents the degeneracy of a given energetic macrostate. For this reason, it is 
common to relate the logarithm of $g(E)$ to the microcanonical entropy,
$S(E)=k_B\ln\,g(E)$, where $k_B$ is the Boltzmann constant. Since all cooperative effects
like phase transitions depend on the interplay of entropy and energy, changes in the monotonic behavior
of $S(E)$ curves indicate the crossover from a macrostate domain in phase space to another one.

The energy function of a system defines a model of it and makes it specific in that typical scales of
lengths and interaction strengths are fixed. If one wishes to study the thermodynamic behavior of 
a class of systems and in order to understand it from a more general perspective, it is desirable to 
express the relevant quantities in a scale-free form. We will proceed so in the following 
for the particularly interesting problem of the adsorption of a flexible polymer to an attractive 
substrate, where different energetic and entropic contributions compete with each 
other~\cite{bj1,eisenriegler1,eisenriegler2,vrbova,singh,bj4,prellberg1,paul1,mbj1}.

\section{Adsorption Model and Estimation of the Contact Density}
In our approach, the polymer with $N$ monomers 
is represented by an interacting self-avoiding walk on a simple-cubic
lattice, i.e., self-crossings are not allowed, but non-bonded monomers being nearest neighbors 
attract each other. The intrinsic energy of the polymer thus only depends on the number of such
nearest-neighbor contacts $n_m$. In addition, a monomer $i$ shall be attracted by a solid, planar 
substrate (defined by $z= 0$), if it resides at a nearest-neighbor position with $z_i=1$. Therefore, the
energy of the polymer--substrate interaction is proportional to the number of monomer--surface
contacts, $n_s$. In units of an irrelevant
overall energy scale, the energy of a polymer conformation is given by~\cite{mbj1}
\begin{equation}
\label{eq:mod}
E^\varepsilon_{n_mn_s}=-n_m-\varepsilon n_s,
\end{equation}
where $\varepsilon$ is the surface attraction strength. It controls the relative energetic impact of 
the formation of monomer--monomer and monomers--substrate contacts.
In order to  
regularize the translational motion of the polymer perpendicular to the substrate, we place an
additional steric wall at $z=L_z$. The polymer is not grafted at the substrate and may move freely in 
the available space between substrate and wall. In the following, we are going to study the 
behavior of a polymer with $N=250$
monomers and set $L_z=300$. 

\begin{figure}[t!]
\centerline{\includegraphics[width=100mm]{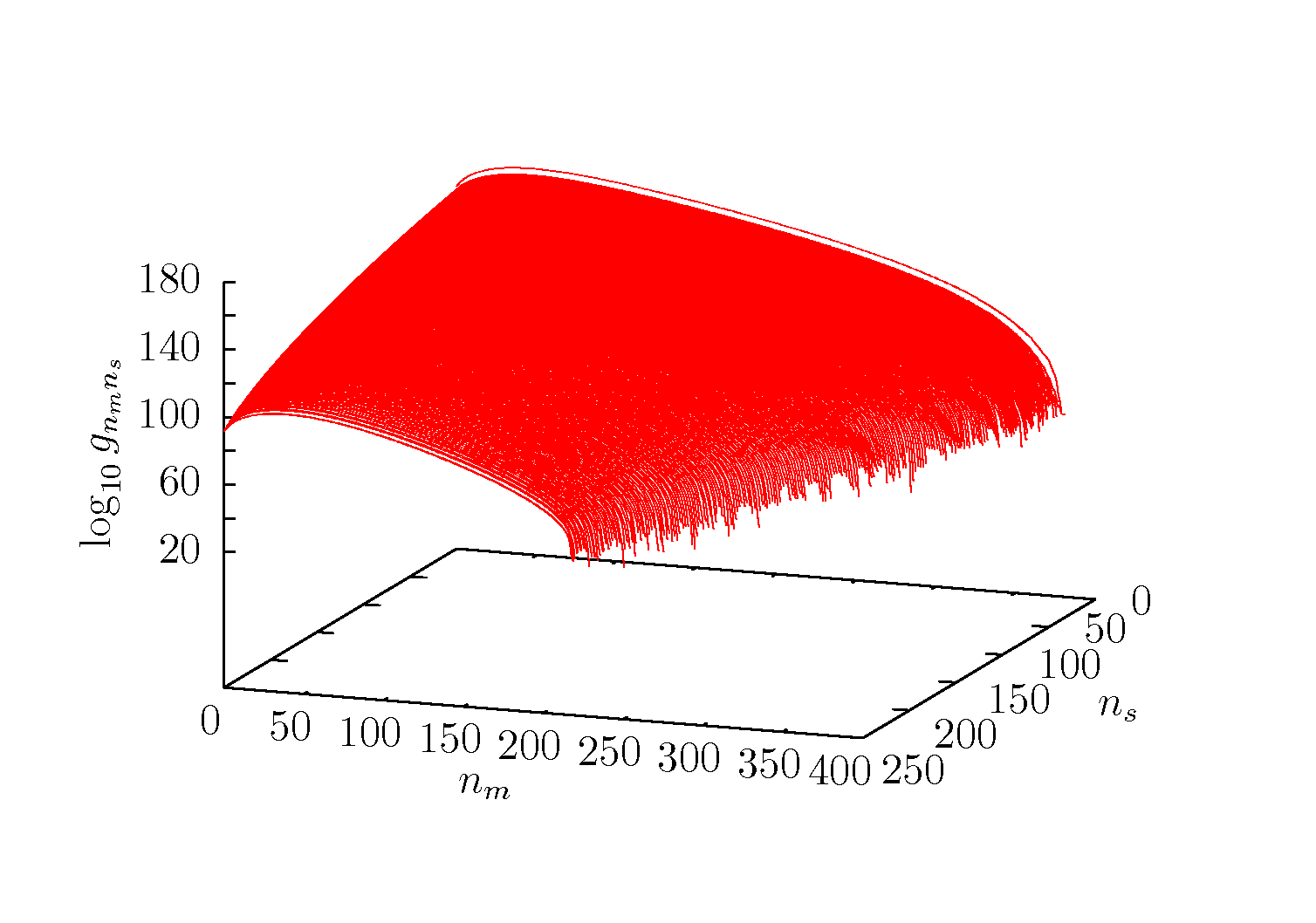}}
\caption{\label{fig:cd} Contact density as a function of the number of monomer--monomer contacts 
$n_m$ and monomer--substrate contacts $n_s$ for a 250mer interacting with a solid substrate.}
\end{figure}

As mentioned above, it would be useful to express the entropy of this system in a scale-free form,
i.e., independently of the system-specific parameter $\varepsilon$. This is possible
by performing computer simulations in a generalized ensemble in which the distribution 
of macrostates in dependence of the contact numbers $n_m$ and $n_s$ is uniform. In analogy to
multicanonical Monte Carlo sampling~\cite{muca1}, the contact-density chain-growth algorithm has been
developed for this purpose~\cite{bj1,bj2,prellberg3}. 
It is a synthesis of Rosenbluth chain growth~\cite{rosen1}, 
population control
by pruning and enriching structure copies~\cite{grass2,hsu1}, and generalized-ensemble sampling.  
The direct result from the simulations is the contact density
$g_{n_mn_s}$. For the 250mer, it is shown in Fig.~\ref{fig:cd}. It is an absolute quantity, i.e.,
the numbers of $g_{n_mn_s}$ in the figure represent the explicit degeneracy of the states. 
Consequently, also the estimate for the 
microcontact entropy $S_{n_mn_s}=k_B\ln g_{n_mn_s}$ is an absolute number.
This is a peculiar property of Rosenbluth chain-growth methods; importance-sampling
Monte Carlo methods can only yield relative entropies. The contact density of the hybrid system 
covers more than 150 orders of magnitude. 
\begin{figure}[b!]
\centerline{\includegraphics[width=90mm]{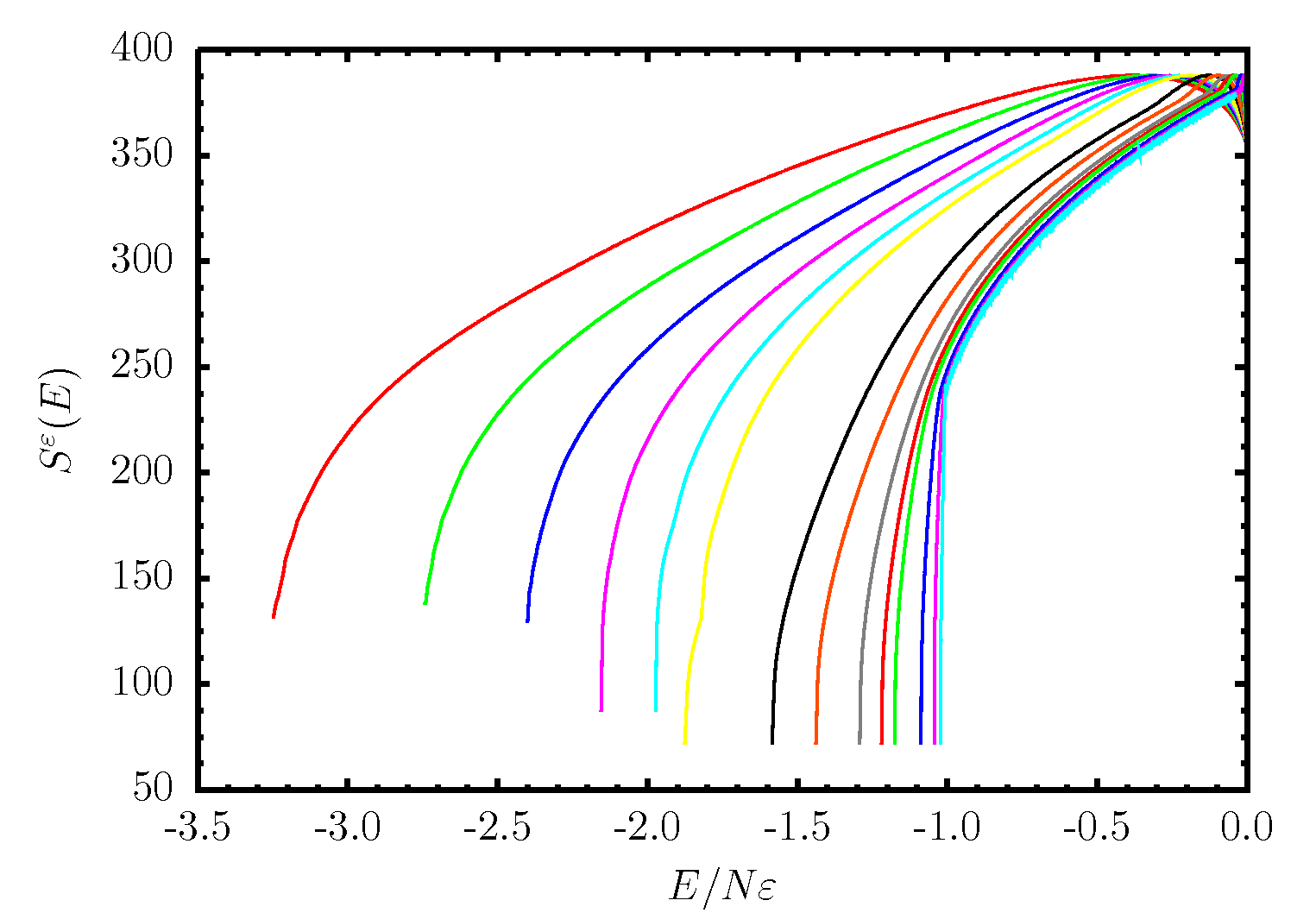}}
\caption{\label{fig:micent} Microcanonical entropies $S^\varepsilon(E)$ of the hybrid system, 
calculated from the 
contact density $g_{n_mn_s}$ for various surface attraction strengths: 
$\varepsilon=0.5,0.6,0.7,0.8,0.9,1.0,1.5,2.0,3.0,4.0,5.0,10.0,20.0,40.0$ (curves from left to right).
}
\end{figure}
\section{Microcanonical Entropy in Dependence of the Surface Attraction Strength}
The advantage of having estimated the contact density becomes obvious if we are interested
in the transition behavior of the hybrid polymer--substrate system for different 
parametrizations of energy scales in the model~(\ref{eq:mod}). Since the microcanonical entropies
can easily be calculated from the contact density,
\begin{equation}
S^\varepsilon(E)=k_B\ln \sum\limits_{n_m,n_s}\delta_{E,E^\varepsilon_{n_mn_s}} g_{n_mn_s},
\end{equation}
where $\delta_{ij}$ is the Kronecker symbol ($1$ if $i=j$, $0$ otherwise), the qualitative differences between the different models are apparent when comparing 
$S^\varepsilon(E)$ for various interaction strengths $\varepsilon$. Several exemplified
entropy curves are shown
in Fig.~\ref{fig:micent} for numerous values of $\varepsilon$ between $0.5$ and $40.0$ ($k_B =1$). 
Note once more that no additional simulations were required to 
obtain these substantially different results. The first interesting observation is that 
for $\varepsilon < 1.0$, the ground-state entropies are larger than for models with 
$\varepsilon \ge 1.0$. The reason is that in the latter case the ground-state conformations
are filmlike (topologically two-dimensional). In this case, it is more favorable for the system
to maximize the number of surface contacts, even at the expense of the reduction of monomer--monomer 
contacts. The ground-state entropy is virtually constant and since
the ground-state energy $E_0$ scales
linearly with $\varepsilon$, $\lim_{\varepsilon\to\infty} E_0/N\varepsilon= -1$. The stepwisely
increasing ground-state entropy for decreasing surface attraction strengths below $\varepsilon=1$
is due to layering effects (``dewetting''). For such model parametrizations, polymer 
ground-state conformations are topologically three-dimensional, i.e., they extend into the space 
direction perpendicular to the substrate~\cite{bj4}. Monomer--monomer contacts are energetically more favorable
than surface contacts. The ground-state entropy increases, because the conformational entropy of a
compact three-dimensional droplet is larger than for a compact film (the number of possible conformations 
is much larger in three dimensions).

The entropy curves also reveal all underlying informations regarding the conformational transitions
such as, e.g., the adsorption transition. For small systems, it is a first-order-like transition
characterized by phase coexistence. This can be investigated best by a microcanonical 
analysis~\cite{gross1,jbj1,mjb1}. 
In the high-energy regime, there is a convex region which causes a bimodal energy distribution at the
transition temperature. Phases of adsorbed and desorbed are separated by a free--energy barrier
which can be traced back to surface effects in the adsorption process. These effects disappear in the
thermodynamic limit, where the adsorption transition is a continuous thermodynamic phase transition.
\begin{figure}
\centerline{\includegraphics[width=90mm]{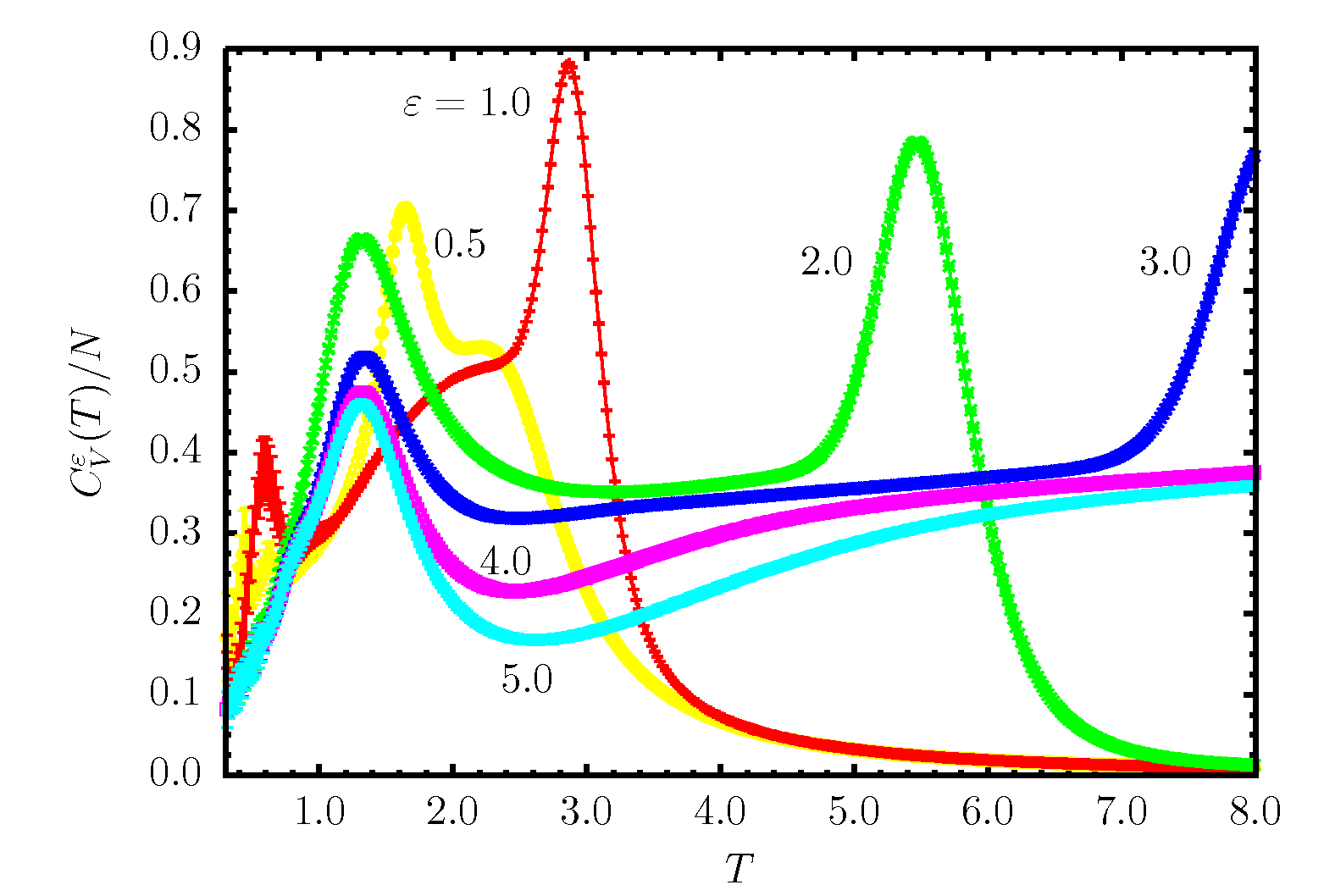}}
\caption{\label{fig:cv} 
Specific heat as a function of temperature for various values of $\varepsilon$.
}
\end{figure}

From the density of states $g^\varepsilon(E)=\exp[S^\varepsilon(E)/k_B]$, the canonical expectation
value of the energy is calculated as 
\begin{equation}
\langle E\rangle^\varepsilon=\frac{\sum_E E g^\varepsilon(E)e^{-E/k_BT}}{\sum_E g^\varepsilon(E)e^{-E/k_BT}}, 
\end{equation}
where $T$ is the heatbath temperature.
The derivative of the mean energy defines the heat capacity: 
$C^\varepsilon_V(T)=d\langle E\rangle^\varepsilon/dT$.
This quantity is shown in Fig.~\ref{fig:cv} for a number of $\varepsilon$ parameter
values. Although canonical averages tend to smear out relevant
statistical informations for small systems, the peak structure of the specific heat 
can give some insights into the
transition behavior. For $\varepsilon\ge 1.0$, the 
pronounced high-temperature peaks indicate
the adsorption/desorption transition.
Peaks and shoulders below these transition temperatures belong to structure formation 
processes on the substrate (such as the collapse from expanded conformations to very compact,
filmlike structures near $T=1.2$ for $\varepsilon\ge 2.0$). An exception is the curve for
$\varepsilon=0.5$ which exhibits a shoulder above the adsorption transitions near $T=2.3$ which 
signals the collapse transition in the desorption regime (i.e., the well-known $\Theta$ collapse
in solvent).

To conclude, we have shown that the contact density is a particularly helpful quantity for understanding 
structural transitions accompanying the adsorption behavior of a flexible lattice polymer. 
Derived from it, microcanonical and canonical
quantities enable the quantitative analysis of the transitions and allow for the 
identification of transition points.

This work is partially supported by the Umbrella program under Grant No.\ SIM6. Supercomputer time
of the Forschungszentrum J\"ulich under Project Nos.\ jiff39 and jiff43 is acknowledged.

%% The Appendices part is started with the command \appendix;
%% appendix sections are then done as normal sections
%% \appendix

%% \section{}
%% \label{}

%% References
%%
%% Following citation commands can be used in the body text:
%% Usage of \cite is as follows:
%%   \cite{key}         ==>>  [#]
%%   \cite[chap. 2]{key} ==>> [#, chap. 2]
%%

%% References with BibTeX database:

\bibliographystyle{elsarticle-num}
%\bibliography{<your-bib-database>}

%% Authors are advised to use a BibTeX database file for their reference list.
%% The provided style file elsarticle-num.bst formats references in the required Procedia style

%% For references without a BibTeX database:

\end{document}